\documentclass[12pt]{iopart}

\usepackage{graphicx}                       
\usepackage{geometry} 
\usepackage{amssymb, latexsym}

\usepackage{iopams}  
\begin{document}

\title[Flow in tire grooves]{Analysis of the water flow inside tire grooves of a rolling car using r-PIV.}

\author{Damien Cabut$^{1}$, Marc  Michard$^{1}$, Serge Simoens$^{1}$, Loic Mees$^{1}$, Violaine Todoroff$^{2}$, Corentin Hermange$^{2}$ and Yohan Le-Chenadec$^{2}$}

\address{$^{1}$ Univ Lyon, Ecole Centrale de Lyon, INSA Lyon, Universit\'e Claude Bernard Lyon I, CNRS, LMFA, UMR 5509, 36 Avenue Guy de Collongue, F-69134, ECULLY, France.\\
$^{2}$ Manufacture Fran\c{c}aise des Pneumatiques Michelin, Clermont-Ferrand, France}
\ead{damien.cabut@ec-lyon.fr}
\vspace{10pt}
\begin{indented}
\item[]November 2020
\end{indented}

\begin{abstract}

In order to better understand the hydroplaning phenomenon, local velocity measurements of water flow are performed inside the tire grooves of a real car rolling through a water puddle. Velocity fields are obtained by combining refraction Particle Image Velocimetry (r-PIV) illumination, seeding fluorescent particles and either the classical 2D2C or a 2D3C stereoscopic recording arrangements. The presence of some bubble columns inside the grooves is highlighted by separate visualisation using fluorescent contrast technique evidencing two phase flow characteristics.
A simple predictive model is proposed supporting the r-PIV analysis. It provides useful information to adjust the focusing distance and to understand the effect of the bubble column presence on the recorded r-PIV images, especially for the seeding particles located in the upper part of the grooves, as fluorescent light is attenuated by the bubbles. Also, the predictions provided by the model are compatible with the measurements. The velocity fields inside the grooves are analysed using ensemble averaging performed over a set of independent snapshots, recorded with the same operating parameters. The variability of the longitudinal velocity distribution measured in a groove for several independent runs is explained by different mechanisms, like the random position of fluorescent seeding particles at various height of the groove, the hydrodynamic interactions between longitudinal and transverse grooves, and the random location of the transverse grooves from one run to an other. Three velocity components in cross-sections of the longitudinal grooves are obtained using the stereoscopic arrangement. They are compatible with the presence of some longitudinal vortices also assumed in the litterature.  The number of vortices is shown to be dependent on the aspect ratio characterising a groove's rectangular cross-section. We demonstrate, from measurements performed for several car velocities, that the velocity distribution inside longitudinal grooves shows self-similarity when using specific dimensionless length scale and velocity. Hydrodynamic interactions between longitudinal and transverse grooves are discussed on the basis of a mass budget; a fluid/structure interaction mechanism is proposed in order to correlate the overall direction of the flow in a transverse groove with its location inside the contact zone. Finally some physical mechanisms are suggested for the birth of longitudinal vortices. 
\end{abstract}

%
\vspace{2pc}

\section{Introduction}

Hydroplaning of tires is a well-known phenomenon which may occur when a tire rolls over wet or flooded roads. This mechanism is due to hydrodynamic pressure building up in front of a tire (Horne and al. 1963\cite{horne1963phenomena}). The application of this pressure on the tire surface in front of the contact patch area generates a lift force that carries part of the load initially carried by the tire. The load carried by the tire is therefore decreased, as well as the length of the contact patch area. Eventually, the lift force can get as high as the load of the vehicle and then the tire loses all contact with the ground leading to a full hydroplaning situation.

Hydroplaning depends on several parameters such as vehicle speed, water film depth, road macro-texture and tires thread pattern. Hydroplaning reduction essentially relies on tire tread design optimization to drain as much water as possible, without loosing performance requirement in terms of security and road adherence. This requires to understand how the water flows inside the tire grooves, to identify the involving physical processes and the key parameters of the problem. In this paper, a dedicated optical method, based on Particle Image Velocimetry is presented and applied on a test track to measure the velocity fields inside the tire grooves of a car rolling through a water puddle. The paper is organized as follows. A literature review on the impact of tread design on hydroplaning and velocity measurements inside grooves is presented in Sec. II. The experimental setup is described in Sec. III, with focus on the specific illumination method imposed by the test track constraints. A preliminary analysis of the raw images recorded for both visualization and PIV measurements is presented in Sec. IV, to highlight the occurrence of two-phase flow in the grooves. Sec. V proposes a simple optical model to estimate bias errors in r-PIV measurements and to account for the presence of bubbles. In Sec. VI velocity fields inside tire grooves are discussed, with focus on the local flow variability and the influence of the location of the transverse grooves inside the contact patch area between ground and tire. In Sec. VII physical mechanisms are proposed to explain the main features of the water flow inside both longitudinal and transverse grooves. Finally Sec. VIII provides some conclusions and perspectives of this work.

\section{State of the art}
\subsection{Impact of tread design on wet braking.}

Hydroplaning process depends on several parameters such as vehicle speed, water depth, road macrotexture or tread design. The tread design impact on the wet braking performances of tires is known since a long time through different studies (Maycock 1965\cite{maycock1965second} and Meades 1967\cite{meades1967braking}). These studies demonstrated that a change of tread design has an important global impact especially from ribbed to smooth tread patterns that is the classical way of life as new to worn tire state.

The understanding of how a specific tread pattern can affect locally the tire adherence mechanisms is studied in the literature through various strategies. Fwa and al. 2009\cite{fwa2009effectiveness} do numerical fluid/structure interaction simulations to determine the critical speed at which hydroplaning is initiated (speed at which no contact remains between the tire and the road) and this for various tread designs. Different parameters were tested, mainly the direction, compared to circumferential one, of the grooves. All proposed designs and grove directions were evaluated for various water depths (from 1 mm to 10 mm height). For all combinations, the authors demonstrated that a change of tread pattern can result in a change of hydroplaning speed up to 30 km/h. Nevertheless the paper did not provided any local information on flow structures and velocity, inside the grooves.

The influence of tread pattern on hydroplaning process as well as the trade-off with several performances has also been extensively studied by Wies and al. 2009\cite{wies2009influence}. The authors tested tires with different compounds as well as tread patterns as a function of the variation of total void volume or longitudinal and lateral void spatial distributions. They demonstrated that available void volume was highly significant to explain their hydroplaning test results.  At iso-void volume, the void spatial orientation was also an important parameter and the longitudinal void orientation appeared to be more relevant than the lateral one for this process. 

An experimental study done by Matilainen et al. 2015\cite{matilainen2015tyre} is focused on the loss of the tire surface contact with the road, as a function of new and worn tread depths. Three-axial accelerometers were used to monitor the length of the contact patch in dry and wet conditions. The authors demonstrated that the tire with higher tread depths kept a larger contact patch area with the ground than the tire with worn tread pattern which more generally highlights the impact of the tread pattern on the hydroplaning mechanism.

More recently, Lower et al. 2020\cite{lower2020dynamic} compared the fluid pressure in the contact patch for different tires with various tread patterns on a wet drum test bench. The fluid pressure is shown to be strongly dependent on tread pattern. 

The literature related to local velocity measurements of the flow in the vicinity of the contact patch area using optical techniques is very sparse. To our knowledge, the only published work is the one by Suzuki and Fukijama in 2001 \cite{suzuki2001improvement}. This experimental investigation took place on a test facility in which a glass plate, covered with water, is embedded into the road and allows visualization of the tire/plate contact patch zone in true rolling conditions. The authors added millet seeds in the water as tracers and recorded images of the seed trajectories, with a high speed camera operating at a rate of 500 Hz. The exposure time (670 $\mu m$ )is large enough to record portions of seeding particles trajectory whose length and direction is directly related to the particle velocity while small enough to get almost time resolved and local information. The authors thus obtained instantaneous velocity field in the water bank, in front of the contact patch area. They exploited these results to optimize the tread pattern of a tire, by aligning the groove directions with the water flow. The optimized tire demonstrated a better cornering performance on a wet drum test bench at high speed which is seen to be the result of an improved tread pattern design for hydroplaning. Susuki\textit{et al} \cite{suzuki2001improvement} show that visualization of the contact patch area, flow pressure and velocity measurements in front of the tire constitute an effective approach to optimize the groove geometries and to improve tire hydroplaning performances. Nevertheless, the experimental set-up suffers from several limitations. The diameter of the millet grain seeds lies in the range 1 mm to 1.5 mm for a specific gravity of 1.2. Their ability to follow the flow variations is then questionable while not fully discussed in the paper; The velocity fields are obtained using a rather archaic method based on single image with long time exposure, whose performance is far bellow the standards of nowadays velocity field measurement techniques. Moreover, the contrast of seed particle trajectories is poor and no velocity information can be obtained inside the grooves where additional difficulties arise, due to the presence of bubbles.     
  
Considering these limitations, further progress in tread pattern design improvement relies on more detailed and local measurements, including measurements inside the grooves. It is also crucial to take into account the actual complexity of the flow. In particular, the two phase flow occurring in the tire grooves obviously affect the water flow and must be observed and analysed. An improved approach is thus suitable to identify in details the zones where water does not flow freely, to solve the associated draining limitations and to get better hydroplaning performance. In this paper, a dedicated measurement technique, derived from Particle Image Velocimetry is developed and applied to some test on real In Situ wet track. The technique, called r-PIV (for refraction-PIV), is characterized by a specific illumination, based on a laser sheet refraction (unlike standard PIV based on direct illumination), from the ground window to the water flow, close to the total reflection angle.  

\subsection{PIV adaptation for tire related water flows.}

Particle Image Velocimetry (PIV) is a well extended technique widely used in fluid mechanics to measure instantaneous velocity fields in a plane (Adrian and Yao 1985 \cite{adrian1985pulsed}, Adrian 1991 \cite{adrian1991particle}). For macroscopic flow measurements, a laser sheet generally illuminates the particles, seeding the flow, in a 2D thin section of the flow. Particle images are recorded perpendicular to the laser sheet. The camera lens and the working distance are selected to meet with the desired field of view and associated requirements. In the most usual situations, the depth of view is larger than the laser sheet and the depth of the measurement volume is controlled by the laser sheet thickness. For stereoscopic PIV, two synchronous cameras are used. The measurement volume is still a 2D but thick section as defined by the laser sheet of enlarged thickness. The stereoscopic arrangements allow a 3-component velocity field measurements in a plane. The two camera optical axis are not perpendicular to the light sheet and Scheimpflug arrangements are used on both camera to get focused images in the whole 2D flow section illuminated by the laser sheet.
These standard PIV set-up can not be applied to the present measurements. The track is equipped with a transparent window embedded in the road and which is the only optical access to illuminate and visualize the flow under the tire. 

Using only a single optical access for illumination and image recording is a rather classical situation in microscopic PIV. These measurements are generally performed using a volume illumination instead of a planar laser sheet one. A required quasi-two dimensional measurement is obtained thanks to the small depth of focus of microscopic lenses (Meinhart and al. 2000 \cite{meinhart2000volume}, Vetrano et al. 2008\cite{vetrano2008applications} and Olsen and Adrian 2000\cite{olsen2000out}). The volume illumination method is currently used for flows whose characteristic scales are microscopic and for working distances of few millimeters. Such a strategy is not directly applicable in the present case, having in mind that typical dimensions of the field of view of about $20$ cm x $20$ cm are required for the whole present tire analysis, as well as a working distance about 1 meter. Another difficulty raised by volume illumination is the reduction of signal to noise ratio due to light reflections, especially in confined spaces. These difficulties can be solved by using fluorescent seeding flow particles and filtering the laser light to record only the fluorescent particle images.

In this paper, the r-PIV method is proposed to perform velocity field measurements for the water flow, on a test track, under the tire of a car rolling through a water puddle. The r-PIV method combines both fluorescent particles and a special illumination that bypasses the single optical access constraint. This characteristic illumination is based on the refraction of a light sheet at the interface between the window and the water flow. The resulting illumination is not 2D and the depth of focus is not small enough to limit the measurement volume to a thin flow section. In such a situation, contrast of particle images depends on their positions, with respect to the non uniform light distribution and to the best focus plane defined by the lens. The r-PIV arrangement is described and studied by Cabut et al \cite{cabut2020refraction}). The technique was compared to standard PIV measurements on an academic channel flow allowing both r-PIV illumination and the usual direct planar sheet illumination. The bottom wall of the channel is a prismatic window similar to the one embedded on the floor of the test track. One of the main issue of this a priori validation of the r-PIV method is the spreading of the light sheet thickness along its propagation and whose value can be of the same order than the depth of focus in the area of interest. Consequently the location and the accuracy of the effective velocity measurements in the liquid phase appeared to be a complex function of the emitting and receiving optical properties. Cabut et al. 2019 \cite{cabut2019particle} successfully used the r-PIV method for the measurements of the local water flow velocity in front of a real tire rolling in a water puddle covering a smooth flat ground (road). Different characteristic zones of the flow, have been identified : the so-called water bank (WB) in front of the median part of the tire, and the shoulder (S) zone in the sides of front part of the tire. Characteristic velocities defined in both zones appear to scale linearly with the car speed for low and medium car regimes, while some departure with linearity occurs for the highest speeds approaching the situation of full hydroplaning. Nevertheless, no measurements in the tire groove water flow were obtained.

\section{Experimental set-up.}
\label{ExperimentalSU}
\subsection{Facility and set-up.}
\label{Facility}

\begin{figure*}
    \centering
    \includegraphics[width=0.8\textwidth]{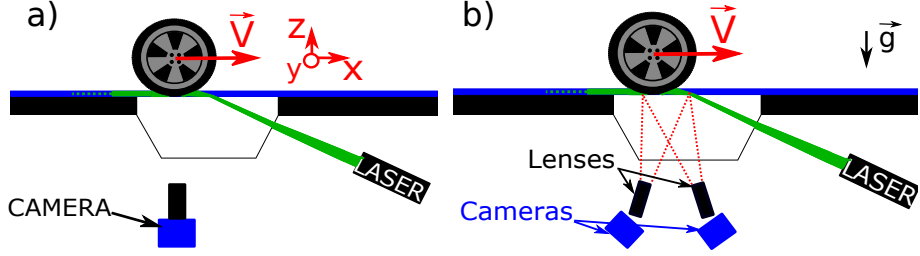}
    \caption{Scheme of the illumination method for a) a single camera b) a stereoscopic mounting.}
    \label{fig1}
\end{figure*}

The test facility used for the present study is located at the Michelin Technology Center in Ladoux (France). The facility consists in a straight track equipped with a transparent and smooth window embedded into the ground. A large area of the track including the window zone can be flooded with a predefined water film depth thanks to an hydraulic closed loop device including a large tank and a regulation system. This facility is used either for direct visualizations of the contact patch area using fluorescence contrast methods (Todoroff et al. 2018\cite{todoroff2018mechanisms}), or for velocity measurements using r-PIV. For safety considerations all devices including hydraulic loop, optical components and command devices are located in a laboratory below the track, as depicted in Fig.\ref{fig1}.

Fig.\ref{fig1} a) and Fig.\ref{fig1} b) display the optical arrangements used for 2D2C (2 dimensions, 2 velocity components) r-PIV measurements and 2D3C (2 dimensions, 3 velocity components) stereoscopic r-PIV measurements. For both configurations, the refraction based illumination method  proposed in Cabut et al. 2020\cite{cabut2020refraction} is used. A laser light sheet enters the prismatic glass window at normal incidence. It reaches the upper face of the window at the interface between glass and water with an angle close to the total reflection critical angle. The light sheet is thus refracted from the glass window into the water film nearly an horizontal outgoing direction; emitting and receiving optics are described in more details in the following Section. 

For direct visualizations of the contact patch zone under rolling condition, the water is mixed with fluorescent dye in order to improve the contrast between the contact patch zone and the flowing water (Todoroff et al.2018\cite{todoroff2018mechanisms}). For such method, the area of interest inside the water puddle is illuminated with continuous light source and images are recorded with an high speed camera.

In the present paper, we use the coordinate system depicted in Fig.\ref{fig1}. The horizontal streamwise direction x is parallel to the car velocity vector, y is the horizontal spanwise direction, and z denotes the vertical axis perpendicular to the flat ground. The velocity components in each direction are respectively U, V and W. Coordinates and velocities can be normalized by $h$ (height of the groove) and $V_{0}$ (vehicle rolling speed) respectively. Their normalized values are denoted ($x^{*}$, $y^{*}$, $z^{*}$) and ($U^{*}$, $V^{*}$, $W^{*}$). The exact position of the tire on the transparent window undergoes some jitter from one run to an other; therefore the origin O of the coordinate system needs to be defined for each run thanks to a special digital image procedure described in the next section.

\section*{Figures}

\subsection{Optical characteristics for r-PIV.}
\label{OpticalChar}
\subsubsection{Emission.}
\label{subsection_emission}

The laser used for these measurements is a frequency doubled, dual cavity Nd:Yag pulsed laser producing short pulses of wavelength $\lambda=532$ nm. The light sheet is generated with an optical arrangement composed of a spherical convergent lens of focal length $200 mm$ and a cylindrical divergent lens of focal length $-40 mm$ to enlarge the laser beam in the y spanwise direction. The window with a prismatic shape has smooth horizontal and inclined faces, and the angle between its inclined and horizontal faces is adjusted around the critical angle for full refraction at the solid/water interface. Cabut et al.\cite{cabut2020refraction} have shown that the inclination of the light sheet propagating inside the window before the solid/water refraction needs to be finely tuned around the total refraction angle at the solid/water interface. Therefore, the whole emitting optical device is mounted on a rotation stage in order to finely adjust the incident light sheet angle at the solid/water interface with a precision of $0.1^{\circ}$. The light sheet vertical intensity profile in the tire grooves is measured when the car is at rest using the same device as described in Cabut et al.\cite{cabut2020refraction}. The profiles obtained for tire measurements are presented in Cabut et al. 2019 \cite{cabut2019particle} (Fig.7) and appear to be nearly uniform both at inlet and outlet of the longitudinal grooves.

\subsubsection{Seeding.}

In order to increase the signal to noise ratio, spurious light reflections or residual light are avoided by using fluorescent particles whose maximum emission wavelength ($584 nm$) is higher than the one of the incident light source ($532 nm$). The particles density is $\rho_{p}=1.19 kg/m^3$. They are almost spherical and show a nearly Gaussian diameter distribution whose mean value is $d_{p}=35$ $\mu$m and standard deviation is about $5$ $\mu$m. Their Stokes number is defined as $St=\frac{\tau_{p}}{\tau_{f}}$, where $\tau_{p}=\frac{\rho_{p}.{d_{p}}^{2}}{18.\mu}$ is the relaxation time of the particle under viscous drag ($\mu$ representing water viscosity). $\tau_{f}=\frac{L_{c}}{V_{0}}$ is a characteristic time for convective effect and is based on the length $L_{c}$ of the contact patch area between the tire and the ground, and on $V_0$ that is a characteristic velocity of the fluid inside the grooves. With this formulation, the order of magnitude of the Stokes number in this experiment is $St=0.01$, small enough to consider the particle as a passive tracer of the flow. 
 
\subsubsection{Image recording.}

Images are recorded with one or two double frame sCMOS cameras with $2560 \times 2160$ pixels. Cameras are equipped with a lens of focal length $100$ mm. The size of the field of view is  ($225$x$190$ mm) and is large enough to record simultaneously particle images in the so-called water bank zone, in front of the tire (Cabut et al. 2020\cite{cabut2020refraction}), and also inside the grooves in the contact patch area, with sufficient spatial resolution. Velocity measurements and flow analysis inside this latter zone are the main purpose of the present paper. 
For the stereoscopic arrangement including two cameras the lenses were mounted with a Scheimpflug arrangement in order to optimize the recovery of the focused zone inside the field of view of each camera. In order to remove incident wavelengths which are source of noise, band-pass optical filters, centered at $590$ nm and with a bandwidth $\pm20$ nm, were placed in front of the lenses. The aperture number was $F^{\sharp}=5.6$ for both 2D2C and 2D3C sets of runs. The choice of the working parameters (focal length, aperture, working distance) result from a compromise between spatial resolution requirements, illumination and technical constraints.
The focus adjustment was made using a ruler placed horizontally on the window for the 2D2C runs and using a 3D calibration target placed on the prism for the 2D3C ones. Therefore, the object plane was located at $z=0$ mm for the 2D2C runs and at $z=1.5$ mm for the 2D3C ones due to the 3D calibration plate structure for stereoscopic measurements. The calibration target is composed of white calibrated dots on a black plate with dots at different heights. To calibrate the stereoscopic recordings, a polynomial algorithm calculated the 3rd degree polynomial that associates coordinates on both camera sensors to a location in the 3D recording area (Prasad 2000\cite{prasad2000stereoscopic}). Based on this calibration, the three velocity components (U,V and W) are derived from the 3D displacements of the particles and the time delay between the laser pulses (Arroyo and Greated 1991\cite{arroyo1991stereoscopic}, Van Doorne and Westerweel 2007\cite{van2007measurement}).

\subsubsection{Synchronizations.}

Synchronization between laser pulses and double frame image recording is performed using a commercial software (Davis8) and a programmable timing unit (PTU). Optical sensors, located along the In Situ track, some meters upstream the water puddle, measured the actual car velocity on the dry part of the track. A trigger signal was then generated with a specified time delay using a real-time dedicated processor and sent to the PTU. This time delay is adjusted in order to synchronize laser pulses and cameras' exposure with the tire passing on the transparent window and more particularly in the field of view. A typical value of the time delay between laser pulses is $180 {\mu}s$ for a car rolling at $V_{0}=13.9 m.s^{-1}$. The repetition rate of both pulsed laser and camera (respectively $15$ Hz and $60$Hz) were not high enough to perform time resolved measurements on a single run. Therefore, a single PIV image pair was recorded for each run and the analysis is based on run series.

\subsection{Tire model.}
\label{TM}
\begin{figure*}
    \centering
    \includegraphics[width=0.8\textwidth]{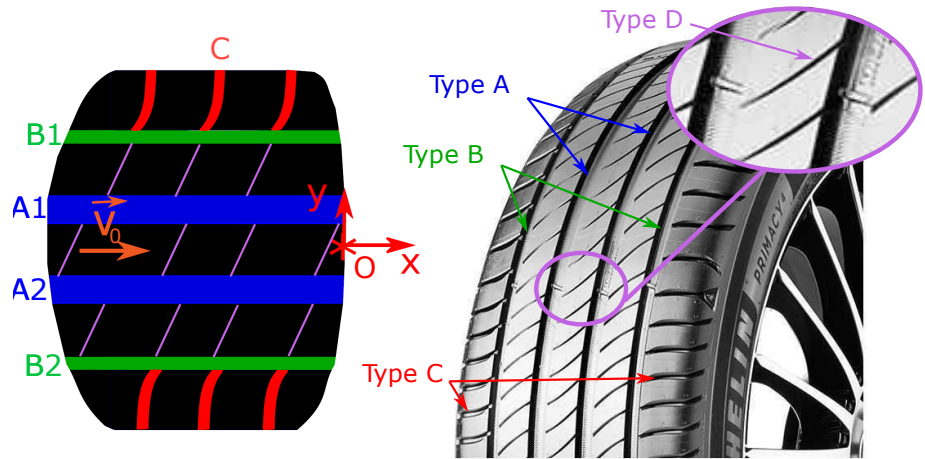}
    \caption{Simplified sketch of the contact patch with different groove types: A (blue), B (green), C (red) and D (purple); the origin 0 of the coordinate system is located at the front edge of the contact patch area.}
    \label{fig3}
\end{figure*}

The present work is focused on the flow inside the grooves of a commercial summer tire (Primacy 4) sketched in Fig.\ref{fig3}. This tire has four types of grooves. The A-type and B-type grooves are longitudinal (oriented in the  direction of motion). The largest ones, hereinafter denoted as A1 and A2, are close to the tire center. The smallest ones, denoted as B1 and B2, are closer to the edges. C-type and D-type grooves are transversal. The tire is in a new state tire, the height $h$ of both A-type and B-type grooves is equal to $6 \ mm$, while their width are respectively 12mm (A-type) and 6mm (B-type). The spacing between the C-type grooves is not strictly constant according to technical issues. Moreover, their cross-section is not rectangular : both height $h$ and width $w$ of a given C-type groove vary at both extremities, with the smallest cross-section near the junction to a B-type longitudinal groove. For such a new tire, the C-type transverse grooves are open-ended at both extremities. Some hydrodynamic interactions can occur either at the junction to the nearest B-type groove or at the open end leading to the external water puddle. Finally, the D-type grooves are very thin transverse grooves linking two A-type or one A-type and one B-type grooves together. Since these grooves are also open at both ends, they are expected to play a role in the overall dynamic of the flow inside longitudinal grooves. However the width of the D-type grooves is too small to contain a sufficient number of seeding particles, to allow velocity measurements due to magnification ratio and optical set-up resolution. Typical values of groove geometrical properties (width, height and shape factor) are summarized in the Table $1$. It is to be noticed that due to the geometry of C-type and D-type grooves, the tire geometry is not strictly symmetric with respect to the vertical mid-plane. Finally, the width $W$ of the tire is 225 mm, and the inflation pressure was fixed at 2.2 bars. 

\section*{Tables :}
\begin{table}[h]
	\centering
	\begin{tabular}{|c|c|c|c|c|c|c|}
		\hline
		Groove type & width w (mm) & height h (mm) & shape factor \\
		\hline
		A & 12 & 6 & 2  \\
		\hline
		B & 6 & 6 & 1  \\
		\hline
		C & 3 - 4 & 6 & 0.5 - 1  \\	    
		\hline		  	    
	\end{tabular}
	\caption{Groove dimensions.}
	\label{table1}
\end{table}

\subsection{Control parameters and measurement procedure}

The two control parameters of the experiment are the water film height $h_0$ and the car velocity $V_0$. In the present work, the water film height was fixed at $h_0=8mm$ which is much higher than the usual water height for realistic wet rolling conditions. Nevertheless this "academic" case is useful to evaluate a processing and a data analysis strategy for well controlled operating conditions before attempting to characterize more realistic cases with smaller water film height and worn tires. The second control parameter, the car velocity $V_0$, varied in the range $8.3 m.s^{-1}$ to $19.4 m.s^{-1}$. The standard deviation of the car velocity as measured over a set of 16 independent runs was typically less than $1\%$, while the standard deviation of the water film height was about $3\%$.

All runs are performed during the night. Note that such an operating mode is useful in order to minimise the spurious ambient light. The car is initially at rest; when the water puddle height is stabilized to $h_0$, the car starts, reaches the specified speed controlled via a speed regulator; the PIV acquisition system is fired when the tire rolls over the desired location on the transparent window, as previously explained. The whole measurement process is repeated over a set of $N=16$ independent runs in order to evaluate the variability of the measurement process and of the flow properties. 

Most of the results are discussed in detail for the case $V_0=13.9$ m/s. The evolution of flow properties inside the grooves as a function of the car speed is detailed in a second step.

\section{Preliminary results : Contrast fluorescence visualizations and r-PIV image processing}
\label{Visu}

\subsection{Contrast fluorescence visualizations : evidence of two-phase flow.}
\label{ContrastFluo}
\begin{figure*}
    \centering
    \includegraphics[width=0.7\textwidth]{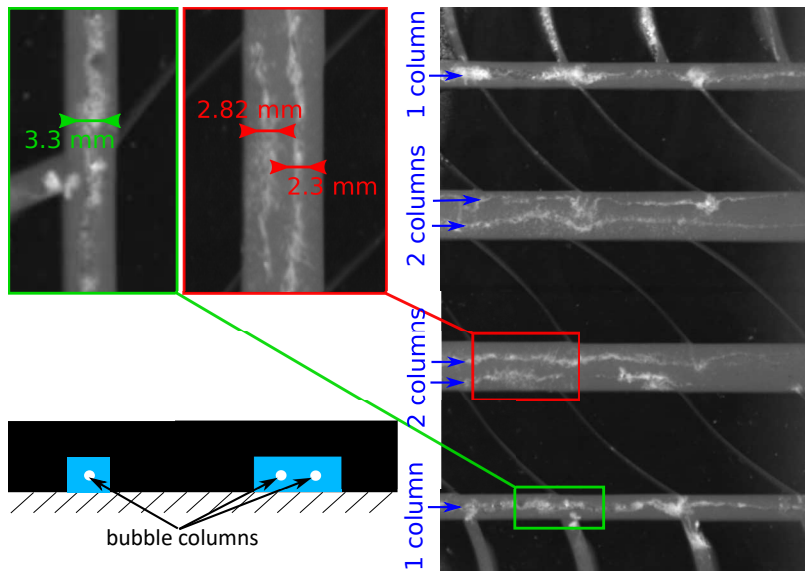}
    \caption{Raw image from visualizations for summer tire with zoom on tire grooves and a scheme of the bubble columns present in the tire grooves.}
    \label{fig4}
\end{figure*}

A sample of raw images of the contact patch area obtained by contrast fluorescence visualization is given in Fig.\ref{fig4}; the view is zoomed around the longitudinal grooves in order to highlight some specific hydrodynamic phenomena. The liquid phase inside grooves appears in grey, while the tire tread in contact with the window is black. The first remarkable feature of the flow inside grooves is the presence of white elongated filaments or columns. This indicates the presence of a gaseous phase (air bubbles or cavity). For the A-type grooves whose geometric aspect ratio is $w/h=2$, two nearly parallel bubble columns are visible, while a single column appears in B-type groove, whose aspect ratio is smaller ($w/h=1$). Nevertheless a detailed examination of the bubble columns in Fig.\ref{fig4} shows some local distortions, mainly inside the largest grooves. These distortions seem to present a more or less periodic behaviour in the streamwise direction, with a wavelength equal to the spacing between two adjacent D-type grooves. Such local disturbances are also visible at the junction between B-type and D-type grooves, as well as at the junction between B-type and C-type grooves. In every cases, these perturbations are probably due either to the impingement of small jets, at the exit of C-type or D-type transverse grooves, on the main flow inside the longitudinal grooves (A-type or B-type) or to some suction effects. Note that the streamwise location of the transverse grooves at the recording time from one run to the another is random. Thus we expect these local interactions to be a possible source of small scale flow variability inside longitudinal groove when performing an ensemble averaging over a set of statistically independent runs.

\subsection{PIV images : definition of a common coordinate system.}

\begin{figure*}
    \centering
    \includegraphics[width=0.6\textwidth]{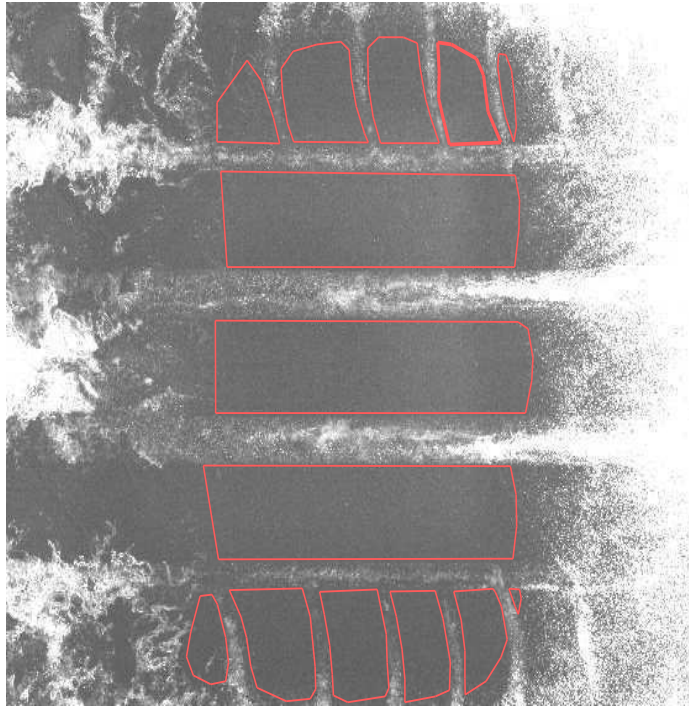}
    \caption{Raw PIV image and mask for summer tire ($V_{0}=13.9 m/s$).}
    \label{fig5}
\end{figure*}

As explained in Sec.\ref{Facility}, the location of the tire in the camera field of view can undergo some jitter from one run to the another. Therefore in order to perform ensemble averaging over the different independent runs, it is necessary to shift instantaneous images (and thus velocity fields) thanks to a coordinate system whose origin is common to the different runs for a fixed set of operating conditions. In the spanwise direction, we choose to shift the coordinates thanks to the position of the side edges of the longitudinal grooves whose location is easily determined from grey levels analysis of raw images. Determining the tire position in the streamwise direction is more difficult. While for direct contrast fluorescence visualisations the edges of the contact patch area between the tire and the ground can be determined using simple image processing tools. This is not the case for the r-PIV images which are less contrasted.
A sample of a raw r-PIV images is shown in Fig.\ref{fig5}. The analysis of the image is performed in front of the central rib which is nearly free of bubble clusters. Near the front edge of the contact patch area, grey level gradients, due to contrast between the rib itself and the ground, are much smaller than the gradients due to fluorescent particles. Moreover the grey level distribution, associated to the seeding particles, shows small scale gradients due to the random location of the particles, while large scale axial gradients are due to an increasing number of seeding particles along $x$. That is correlated to the increase of the clearance height. We note that some motionless isolated seeding particles are trapped between the rib and the ground.

\begin{figure*}
	\centering
	\includegraphics[width=0.6\textwidth]{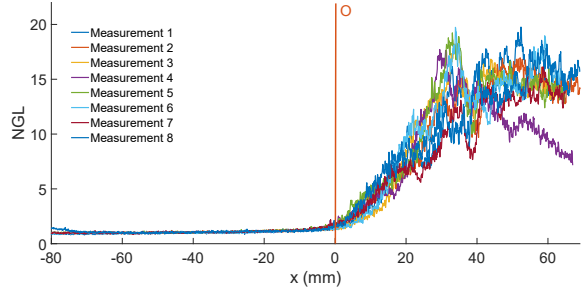}
	\caption{Gray level profiles along the central rib for eight statistically independant runs ($V_{0}=13.9 m/s$).}
	\label{fig6}
\end{figure*}

Therefore we decided to determine the location of the front end of the contact patch area from the large scale variations of grey level in front of the central rib. Small-scale variations are removed by spatially averaging the grey-level in the spanwise direction $y^{*}$ in a narrow band whose width is equal to the width of the central rib. An example of resulting streamwise averaged grey-level profile is shown in Fig.\ref{fig6} for a set of eight independent runs. The length of the bottom out appearing in these profiles corresponds to the length of the contact patch area. The origin O of the coordinate system is then defined as the location $x^{*}$ where such profiles equal to a specified threshold, slightly below the mean grey-level of the bottom out ; in the present experiment, this threshold is fixed to 1.03. This procedure appeared to be satisfactory for all runs in the whole range car velocity $V_0$ explored. A careful examination of such grey level profiles shows that their variability in the immediate neighbourhoods of the front and the leading edge of the bottom out is more likely due to a small global shift in the streamwise $x^{*}$ direction, rather than to a variation of their length.

\subsection{PIV image masking and interrogation areas.}

There is no seeding particles in the contact patch area. Therefore this region must to be masked prior to the PIV analysis based on cross-correlation computation. Despite the test pilot driving skills, the positioning of the transverse grooves undergoes some random jitter from one run to the other (typically around $30$ mm). Furthermore, the geometric properties of the transverse grooves, width and spacing, are not exactly periodic along the azimuthal direction of the tire. Therefore the automatic determination of the contact patch area edges with raw PIV images, especially in the regions of transverse grooves, is a difficult task, beyond the scope of the present work. This task was therefore performed manually by restricting the inner part of the mask to regions where no seeding particles are visible. An example of such a mask is shown with red lines in Fig.\ref{fig5}.  

The PIV analysis is based on cross-correlation between two successive images decomposed into small cells named interrogation windows, with a certain amount of overlapping. Different sizes of interrogation windows have been investigated, from $32$ pixel $\times 32$ pixels down to $12$ pixels $\times 12$ pixels (${\delta}/w=0.09$), with an overlapping factor equal to $50\%$. In D-type grooves, even though some individual fluorescent particles can be identified in raw r-PIV images. The ratio between the window interrogation size ${\delta}$ and the width $w$ of the groove is not large enough to perform the cross-correlation with a sufficient signal/noise ratio. Therefore velocity measurements presented in the following are restricted to A, B and C-type grooves.

\section{Cross-correlation model for bubbly flows}
\label{Met}

In the present experiment, the illumination intensity profile is nearly uniform along the depth of a groove, and the depth of focus is of the same order than the groove depth. Therefore, all individual particles present in the measurement volume along the vertical direction $z^{*}$ contribute to the cross-correlation between images. We are thus facing two questions i) What is, without bubbles and for specified emitting and receiving optic parameters, the effect on velocity measurements of fluorescent particles located at different altitudes $z^{*}$, between the ground and the top of the grooves ? and ii) what is the effect of bubbles on these measurements ? The development of rigorous and exact models to account for these effects is a very difficult and challenging task. We present in this paper a simplified optical model of such last effects, without and with bubbles.

\subsection{Cross-correlation model for one-phase flow.}

\begin{figure*}
	\centering
	\includegraphics[width=0.4\textwidth]{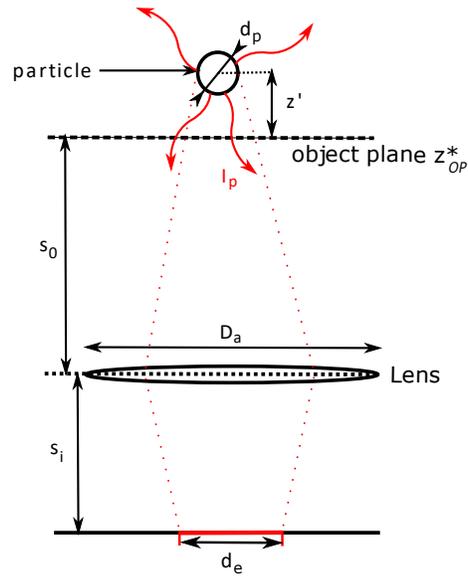}
	\caption{Scheme of the different geometrical and optical parameters of the model.}
	\label{fig7}
\end{figure*}

The starting point is the statistical image cross-correlation model (CCM) developed in Cabut et al.\cite{cabut2020refraction} for a one-phase flow. A sketch of the different optical parameters used in the model is depicted in Fig.\ref{fig7}.

This simplified model predicts the ideal form of the cross-correlation function between images 1 and 2 for a given ensemble of individual fluorescent particles located at different random heights $z_{i}^{*}$, and for a set of specified parameters of the emitting and receiving optics. A set of $P=10$ dimensionless altitudes $z_{i}^{*}$ of particles are first randomly chosen in the range 0 to 1; the value of P is representative of the actual number of particles identified inside an elementary interrogation window inside grooves. The simplified model for the cross-correlation function $R(\textbf{s})$ for an elementary interrogation area is then written as :

\begin{eqnarray}
R(\textbf{s})=\sum_{i=1}^{P} {R}(\textbf{s}|z_{i}^{*})
\label{CCM_1}
\end{eqnarray}

where the cross-correlation function ${R}(\textbf{s}|z_{i}^{*})$ for each individual fluorescent particle $i$ is given by :

\begin{eqnarray}
{R}(\textbf{s}|z_{i}^{*})&=&\frac{A_{i}.{I_{0}}^{2}}{2.{d_{e}}^{2}(z_{i}^{*}).(s_{0}+z_{i}')^{4}}.e^{-4\beta^{2}\frac{({DX}(z_{i}^{*})-s_{x})^{2}+({DY}(z_{i}^{*})-s_{y})^{2}}{2.{d_{e}}^{2}(z_{i}^{*})}}\\
&=&A_{i}.{I_{0}}^{2}.F_{i}..e^{-4\beta^{2}\frac{({DX}(z_{i}^{*})-s_{x})^{2}+({DY}(z_{i}^{*})-s_{y})^{2}}{2.{d_{e}}^{2}(z_{i}^{*})}}\nonumber
\label{CCM_2}
\end{eqnarray}

$\textbf{s}$ is the displacement vector in the correlation plane. $DX$ and $DY$ are respectively the displacement of the image of particle $i$ in $x^{*}$ and $y^{*}$ directions in the sensor plane : $DX(z_{i}^{*})=M.U(z_{i}^{*}).\delta t$ and $DY=M.U(z_{i}^{*}).\delta t$, if $M$ denotes the magnification factor, $\delta t$ the time delay between laser pulses and $I_{0}$ the light sheet intensity which has been previously shown to be independent of $z_{i}^{*}$. $A_{i}$ is a constant factor depending on fixed optical parameters. The weighting factor $F_{i}$ before the exponential term is defined as the focusing function. The effective diameter $d_{e}$ of a particle image seen by the sensor is expressed in the same way as Olsen and Adrian 2000\cite{olsen2000out} :

\begin{eqnarray}
{{d_{e}}}(z_{i})=\sqrt{M^{2}{d_{p}}^{2}+5.95(M+1)^{2}\lambda_{p}^{2}{f^{\sharp}}^{2}+\frac{M^{2}{z}_{i}'^{2}{D_{a}}^{2}}{(s_{0}+{z}_{i}')^{2}}}
\label{diam}
\end{eqnarray}

It is a function of the actual diameter $d_{p}$, the object plane position $z_{OP}^{*}$, the distance ${z}_{i}'^{*}=z_{i}^{*}-z_{OP}^{*}$ between the particle and the object plane, the distance $s_{0}$ between object plane and lens position, the aperture number $f^{\sharp}$ of the lens, the lens diameter $D_{a}$ and the wavelength $\lambda$ of the re-emitted light. 

In addition to the emitting and receiving optical properties, the last input of the CCM model is the actual velocity field $\mathbf{u_{in}}$ to be measured with r-PIV; the model output $\mathbf{u_{out}}$ is the measured velocity, whose components are obtained by dividing the main peak location coordinates of ${R}(\textbf{s})$, in the correlation plane $(s_x,s_y)$, by ${\delta}t$.\\

\subsection{Cross-correlation model for two-phase flows.}

The visualizations shown in Sec.\ref{ContrastFluo} let us presume that the fluorescent light transmission coefficient through the bubble column and for particles located in the upper part of the grooves, is in fact very low. In this section we derive a simple extinction model for a fluorescent particle above such cluster of bubbles.

\begin{figure*}
	\centering
	\includegraphics[width=0.6\textwidth]{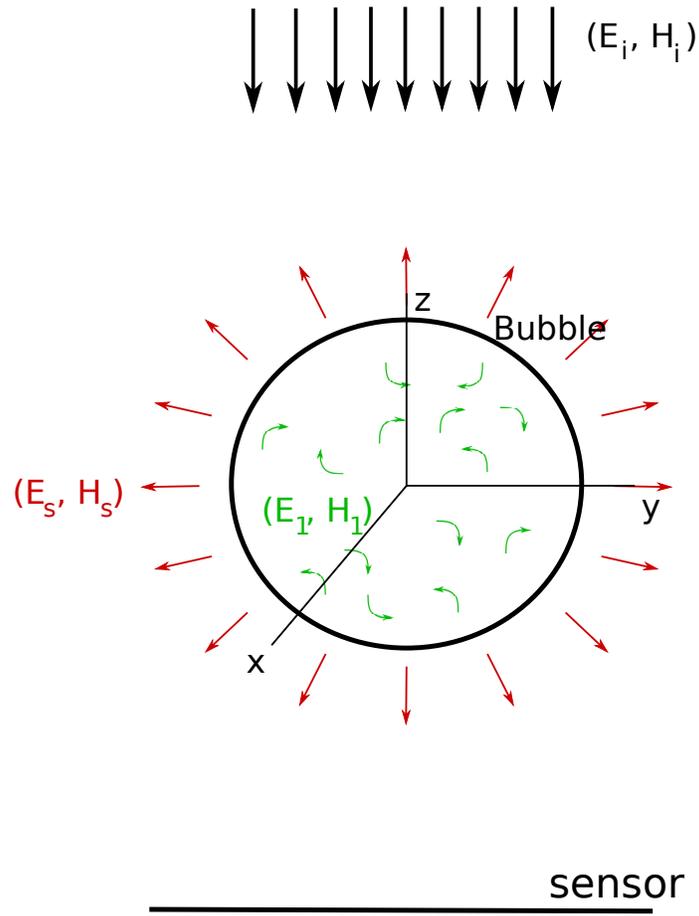}
	\caption{Incident,internal and scattered electromagnetic fields for a spherical bubble.}
	\label{fig8}
\end{figure*}

In a first step we derive the theoretical scattered cross-section for a single spherical bubble submitted to an incident plane wave electromagnetic field ($E_{i}$, $H_{i}$) (Fig.\ref{fig8}). The distance between the fluorescent particle and the bubble is assumed high enough for the incident field to be a plane wave.

\begin{figure*}
	\centering
	\includegraphics[width=0.6\textwidth]{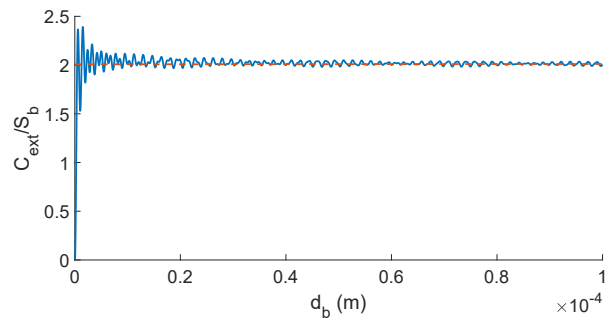}
	\caption{Scattering cross-section normalized by the bubble cross-section as a function of the particle diameter $d_{p}$.}
	\label{fig9}
\end{figure*}

According to Mie's theory, considering each medium (water and air) as local, linear, homogeneous and isotropic, and integrating the Poynting vector of the scattered field at the surface of a sphere surrounding the bubble, Bohren and al 2008\cite{bohren2008absorption} derived the cross-section $C_{ext}$ by resolving the Maxwell equations using suitable boundary conditions. Neglecting absorption for a spherical air bubble, the extinction coefficient $C_{ext}$ can be expressed as a function of the scattered field ($E_{s}$, $H_{s}$). The evolution of the scattering cross-section $C_{ext}$ divided by the physical section $S_{b}$ of the bubble is shown in Fig.\ref{fig9} for a wavelength of $\lambda=584$ nm. For this wavelength, the normalized cross-section converges to value of approximately $C_{s}/S_{b}=2$. For very small bubble diameters (below $10$ $\mu$m), the value of the scattering cross-section oscillates between 0 and 2.5 times the geometrical section.

\begin{figure*}
	\centering
	\includegraphics[width=0.6\textwidth]{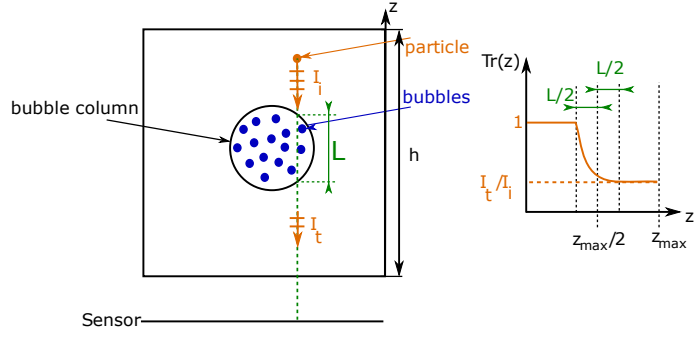}
	\caption{Scheme of a particle above a bubble column (left); evolution of Tr with z (right)}
	\label{fig10}
\end{figure*}

In a second step we consider the case of a slab of spherical bubbles trapped in a cylindrical column as depicted in the Fig.\ref{fig10}. In that case a statistical approach is more suitable. The bubbly medium is statistically considered as an homogeneous medium. More rigorously, at an instant $t$, a particle can have a bubble between itself and the camera, or not, depending on the medium. Statistically it has a probability to be hidden. This is taken into consideration with this approach. Multiple scattering effects (from bubble to bubble) are negligible. The light scattered by a given bubble is not considered as a secondary source of light for another bubble. The attenuation of an incident light propagating through the bubble slab is therefore considered as a continuous change of medium. The attenuation of the light through this medium is calculated based on the Beer-Lambert law as $I_{t}=I_{i}.e^{-\alpha_{ext} .L}$, where $\alpha_{ext}$ is an equivalent extinction coefficient, $L$ the thickness of the medium crossed by the incident light $I_{i}$ and $I_{t}$ is the transmitted intensity.

In the case of a bubble column composed of multiple bubbles, the attenuation coefficient of the bubbly medium is calculated as the averaged attenuation generated by single bubbles multiplied by the density $N_{b}$ of bubbles (in bubble number $/m^{3}$) as $\alpha_{ext}=N_{b}.\overline{C_{ext}}$, where $\overline{C_{ext}}$ is the averaged scattering cross-section of the single bubbles. Finally the value of the transmission coefficient is :
\begin{eqnarray}
Tr=\frac{I_{t}}{I_{i}}=e^{-N_{b}\overline{C_{ext}}.L}
\label{Transm}
\end{eqnarray}
where $L$ is the thickness of the bubble column. The remaining question is to evaluate the value of $\overline{C_{ext}}$.

We assume that the bubble diameter $d_{b}$ is randomly distributed following a normal distribution $p(d_{p})$ centered around the mean value $d_{b}^{(m)}$ and with a standard deviation $\sigma_{b}$ : $p(d_{p})=\frac{1}{\sigma_{b}\sqrt{2\pi}}e^{-\frac{(d_{b}-d_{b}^{(c)})^{2}}{2\sigma_{b}^{2}}}$. The averaged extinction coefficient in the bubble column is thus given by:

\begin{eqnarray}
\overline{C_{ext}}=\int_{0}^{\infty}p(d_{b})C_{ext}(d_{b})dd_{b}
\end{eqnarray}

\begin{figure*}
	\centering
	\includegraphics[width=0.6\textwidth]{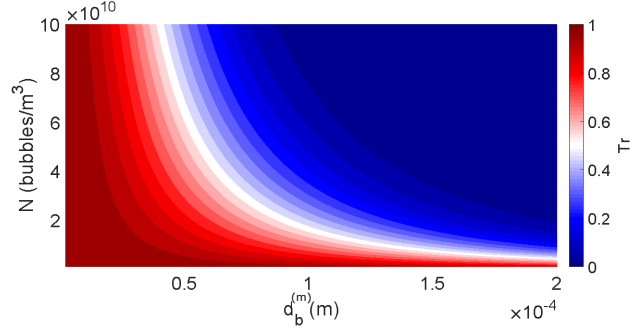}
	\caption{Transmission coefficient Tr as a function of the bubble density N and bubble mean diameter $d_b^{(m)}$.}
	\label{fig11}
\end{figure*}

The thickness $L$ of the bubble column crossed by the light re-emitted by a fluorescent particle is a function of their relative location in the ($y^{*}$,$z^{*}$) plane as sketched in Fig.\ref{fig10}. Fig.\ref{fig11} is a two-dimensional map showing the variations of Tr as a function of the density $N$ of bubbles and of their mean diameter $d_{b}^{(m)}$; the value of $L$ is arbitrarily fixed to $1.5$ mm as evaluated from direct visualizations.

To account for the presence of bubbles in the flow, this transmission coefficient Tr is finally introduced in the CCM model previously developed for r-PIV applied to single phase flow.

\begin{eqnarray}
R(\textbf{s})=\sum_{i=1}^{P} Tr^{2}(z_{i}^{*}).{R}(\textbf{s}|z_{i}^{*})
\label{BCCM_simp1}
\end{eqnarray}

In the following this modified cross-correlation model for two-phase bubbly flow is called BCCM.

\subsection{Application to Taylor-Green vortices.}

In order to test the BCCM in its application to the present case, we need to define, as an input of the model, a presumed velocity field in a cross-section of the groove. As explained previously and according to preliminary results based on contrast fluorescence visualizations, we assume the cross-flow to be described by Taylor-Green vortices, and to be invariant in the streamwise direction $x^{*}$:

\begin{eqnarray}
V(x^{*},y^{*},z^{*})=a.sin(2\pi*y^{*}*h/w)cos(\pi*z^{*})\nonumber\\
W(x,y,z)=-a.cos(2\pi*yx^{*}*h/w)sin(\pi*zx^{*})
\label{Taylor}
\end{eqnarray}

\begin{figure*}
	\centering
	\includegraphics[width=0.6\textwidth]{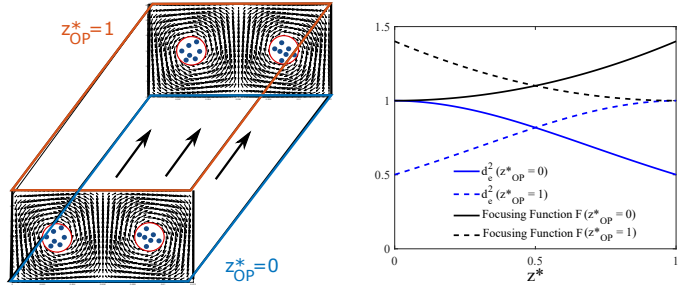}
	\caption{Location of the object plane and effective particle image diameter as a function of z for both configurations.}
	\label{fig12}
\end{figure*}

For an A-type groove with $w/h=2$, the structure of such a pair of counter-rotating vortices is depicted in the Fig.\ref{fig12} (left); for a B-type groove with $w/h=1$, the cross-flow is composed of a single vortex. The characteristic velocity $a$ is an indicator of the strength of the vortices. Such an analytic velocity field is compatible with an impermeability condition on the four side walls of the groove, but not with the actual no-slip condition. Therefore this simplified model doesn't take into account i) the presence of viscous boundary layers near the walls of the grooves, ii) the local disturbances of the flow at the junction between longitudinal and transverse grooves previously highlighted in Sec.\ref{TM} and iii) the local distortion of the Taylor Green vortex velocity induced by the bubble columns.

According to the contrast fluorescence visualizations shown in (Fig.\ref{fig4}), for an A-type groove, we assume the existence of two bubble columns of thickness $L$ equal to $1.5$mm and which are trapped at the center of each presumed vortex (one per vortex). BCCM was run for two different receiving optic configurations, with the object plane located either at the ground ($z_{OP}^{*}=0$)  or at the top of the groove ($z_{OP}^{*}=1$) to demonstrate the effect of bubble column on particle seeding fluorescent light emission. The evolution, with $z^{*}$, of the focusing function $F$ is shown in Fig.\ref{fig12} (right) for the receiving optical parameters given in Sections \ref{ExperimentalSU} and \ref{OpticalChar}. The function is presented normalised by its value at the object plane position $z_{OP}^{*}$. Since the depth $h$ of the groove is much smaller than $s_{0}$, the global shape of the curves is mainly dictated by the variations of $d_{e}$ with $z^{*}$. It appears that the highest departure of the focusing function ($F$) from its maximal value is around $20\%$. Therefore, despite an uniform light sheet intensity profile, for $z_{OP}^{*}=0$ we can expect the properties of the receiving optics to promote, in a statistical manner, the contribution of particles located in the lower part of the groove. 
On the other hand, for $z_{OP}^{*}=1$ the contribution will correspond to the particles located in the upper part of the groove.    

\begin{figure*}
	\centering
	\includegraphics[width=0.6\textwidth]{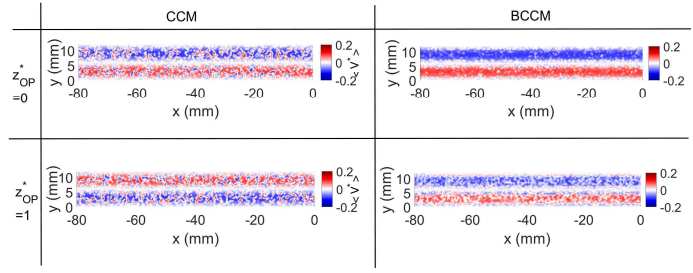}
	\caption{Spanwise velocity map predicted by CCM (left) and BCCM (right), for $z^{*}_{OP}=0$ (top) and $z^{*}_{OP}=1$ (bottom); one single run $N_{im}=1$}
	\label{fig13}
\end{figure*}

\begin{figure*}
	\centering
	\includegraphics[width=0.6\textwidth]{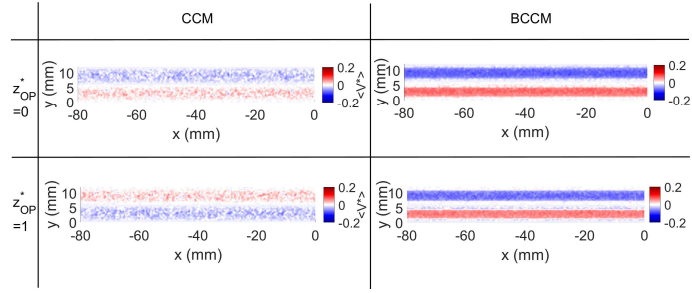}
	\caption{Spanwise velocity map predicted by CCM (left) and BCCM (right), for $z^{*}_{OP}=0$ (top) and $z^{*}_{OP}=1$ (bottom); ensemble averaging over $N_{im}=10$ independant runs}
	\label{fig14}
\end{figure*}

Spanwise velocity component results from both CCM and BCCM predictions obtained with such conditions with vortices (thus with or without bubbles) are shown in Fig.\ref{fig13} and Fig.\ref{fig14}, using respectively $N_{im}=1$ and $N_{im}=10$ image pairs per interrogation window, and for $P=10$ particles per interrogation window. Results are given for both positions of the object plane.

The CCM predictions shown in Fig.\ref{fig13} highlight a small scale intermittent feature of the spanwise velocity component $V^{*}$ due to the random vertical position $z_{i}^{*}$ of the seeding particles. Nevertheless, when the object plane is near the bottom of the groove ($z_{OP}^{*}=0$), according to the decrease with $z^{*}$ of the focusing functions $F_{i}$, shown in Fig.\ref{fig12}, the contribution to $R(\textbf{s})$ of seeding particles near the ground is dominant. The predicted velocity profile $V^{*}(y^{*})$ is compatible with the actual flow directed towards the side wall, near the ground as depicted in Fig.\ref{fig12}a). On the other hand, if the object plane is near the top of the groove ($z_{OP}^{*}=1$), the particles near the top of the groove are most dominant in the expression of $R(\textbf{s})$. Therefore if two-phase flow phenomena were not present inside grooves, the measured horizontal spanwise velocity component $V^{*}$ should be very sensitive to the altitude $z_{OP}^{*}$ of the object plane.

\begin{figure*}
	\centering
	\includegraphics[width=0.6\textwidth]{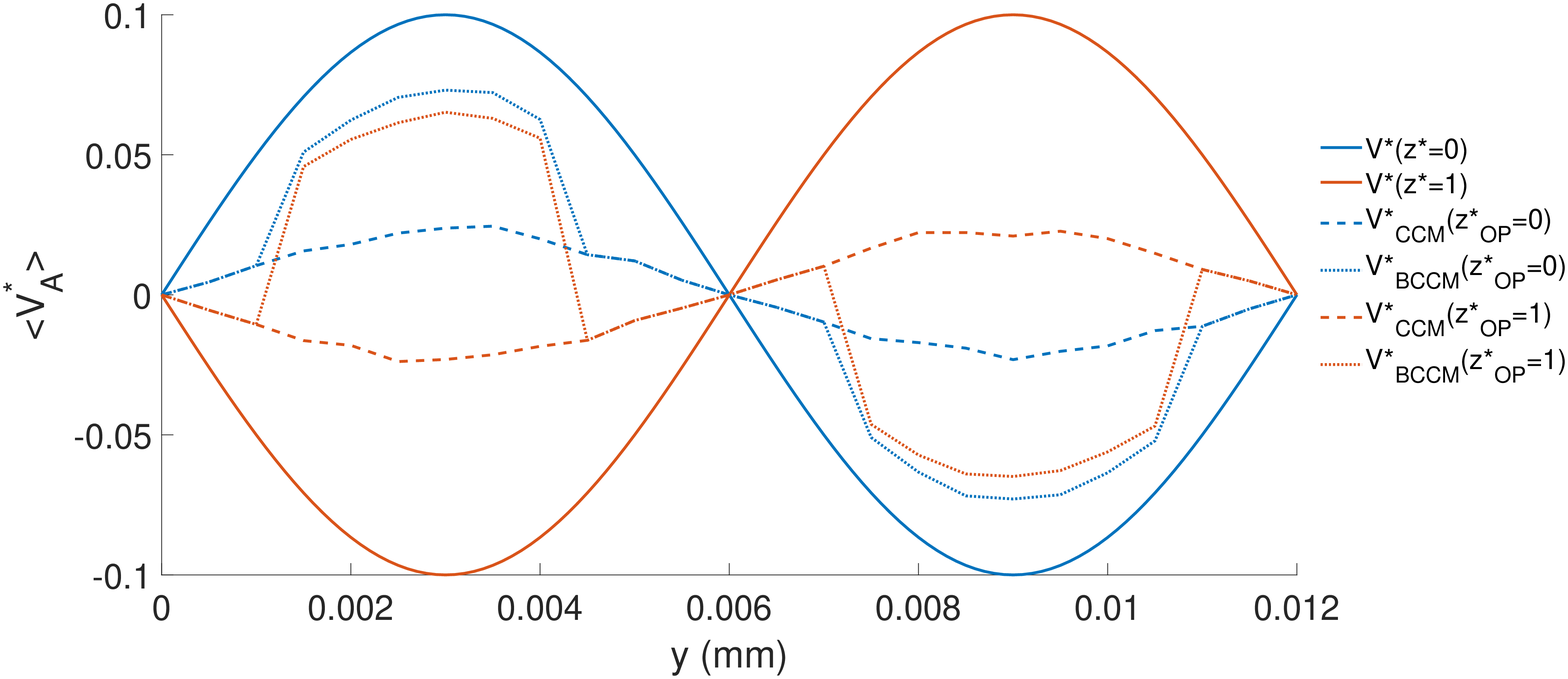}
	\caption{Spatially averaged velocity profiles (actual, predicted by CCM and BCCM) for both object plane positions $z^{*}_{OP}=0$ (blue) and $z^{*}_{OP}=1$ (red).}
	\label{fig15}
\end{figure*}

The BCCM predictions are completely different when the masking effect of bubbles is taken into account. The maps shown on the right of Fig.\ref{fig13} show that there is no reversal of the $V^{*}(y^{*})$ spanwise velocity component when the object plane is moved from the bottom to the top of the groove. Whatever is the position of the object plane, the measured transverse velocity is representative of the actual velocity near the ground, and is not compatible with the actual one at the top of the groove. When the number of image pairs $N_{im}$ is increased, these conclusions are enforced (Fig.\ref{fig14}) and more visible in Fig.\ref{fig13} when the previous spanwise velocity $V^{*}$ component is spatially averaged along the homogeneous $x^{*}$ direction, as done in order to remove small scale fluctuations, shown in Fig.\ref{fig15} and due to the random altitude $z_{i}^{*}$ of the simulated particles.

In conclusion, in order to avoid optical measurement biases, due to uncontrolled local properties of the bubble columns, and to increase the focusing of the particles in the most visible part of the flow (below bubble columns), we choose to place the object plane in the lower part of the groove, near the ground level at $z_{OP}^{*}=0$ for standard 2D2C measurements, or $z_{OP}^{*}=0.25$ for stereoscopic measurements). The different measured velocity components are representative of the flow below the bubble columns.

\section{Flow inside A and B-type grooves.}

\subsection{Instantaneous velocity map - Boundary layers - turbulence.}
\begin{figure*}
	\centering
	\includegraphics[width=0.6\textwidth]{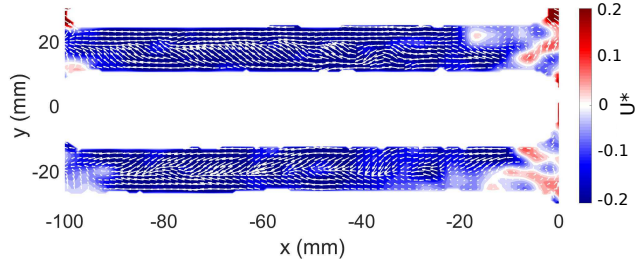}
	\caption{Instantaneous velocity field inside A an B grooves ($V_{0}=13.9 m/s$).}
	\label{fig16}
\end{figure*}

An example of an instantaneous vector map of the velocity measured inside A and B grooves is depicted in the Fig.\ref{fig16}. The associated background color map shows the streamwise velocity component $U^{*}(x,y)$ which is, along both longitudinal groove types, negative, opposite to the car velocity direction $V_{0}$.
A detailed analysis of the instantaneous velocity map shows local inclinations of the velocity vectors.

As seen on this velocity map the longitudinal velocity can locally reach values down to $-0.1V_0$. The Reynolds numbers $Re_{w}$,  based on this characteristic velocity and the width $w$ of the grooves, is around $3000$ for $V_{0} = 50km/h$. Note that due to the unsteady nature of both geometry and flow in a fixed reference frame (linked to the ground), the value of this Reynolds number is not useful for a discussion related to boundary layers. The part of the groove deformable solid walls which are located inside the contact patch area (side walls and upper wall) can by definition be considered to have a zero velocity. With the adherence condition at the lower wall (transparent window), despite the car motion, the flow inside the groove can be assimilated to the flow in a channel with adherence conditions on the four walls and in the fixed reference frame; therefore the velocity of the liquid phase should be zero on the boundaries. Nevertheless transverse average measurements shown on Fig.\ref{fig16} for both groove types don't evidence such an adherence condition, neither for A nor B-types. The explanation is that the boundary layer thickness is too small compared to the size of the PIV analysis interrogation window, and viscous effects are not resolved.   

The second question is relative to the laminar or turbulent properties of the flow. According to the previous discussion related to the boundary layers, even if turbulence is probably produced near the side solid walls, its typical length scales are not high enough to be spatially resolved in these areas. Nevertheless we can imagine other production mechanisms of turbulence for the present flow, even far from side walls. Visualizations shown Fig.\ref{fig4} let suggest strong interactions between small scale jets emanating from an open end of a transverse groove and impinging the adjacent connected longitudinal groove. Moreover the same visualizations show distortions of the bubble column; it is likely that these columns can show some instability mechanisms and locally break into small scale random motion. 

\subsection{Flow through the longitudinal grooves.}
\label{Long}

\begin{figure*}
	\centering
	\includegraphics[width=0.6\textwidth]{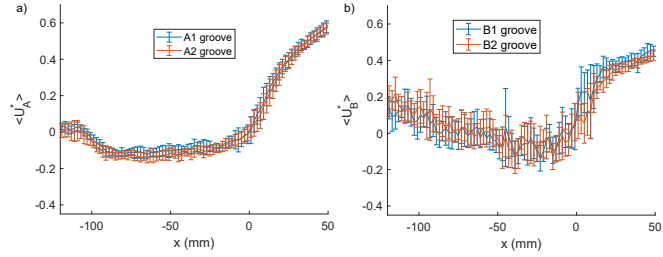}
	\caption{Ensemble averaged velocity profiles $<U_{A}^{*}(x)>$ (left) and $<U_{B}^{*}(x)>$ (right); x=0 mm denotes the front edge of the contact zone}
	\label{fig17}
\end{figure*}

In this section, we are focused, for $V_{0}=50 km/h$ case, on the velocity component $U^{*}$ parallel to the longitudinal groove direction. 
For every longitudinal grooves A1, A2, B1 and B2 the streamwise measured velocity $U^{*}(x,y)$ is first spatially averaged in the $y^{*}$ spanwise direction as: $U_{A}^{*}(x)=\frac{1}{w}\int U^{*}(x,y)dy$ for A-Type grooves, where $w$ is the width of the groove; the set of statistically independent profiles $U_{A}^{*}(x)$ are then ensemble averaged. The same values are defined concerning the B-type grooves as $U_{B}^{*}$. The Fig.\ref{fig17} shows the resulting profiles $<U_{A}^{*}(x)>$ for A-type grooves (left) and $<U_{B}^{*}(x)>$ for B-type grooves (right) respectively; solid lines are the mean values while the variability of measurements is characterized, for every $x$ location, by vertical bars with an amplitude equal to +/- the associated standard deviation around the mean. The contact zone is delimited between approximately $x=-90$ mm and its front edge at $x=0$ mm. 

For both longitudinal groove types, in front of the contact patch area, the fluid velocity $<U_{A}^{*}(x)>$ decreases from positive values, near $x=50$ mm where water is pushed away from the tire, down to negative values at the inlet ($x=0$ mm) of the contact zone. Indeed we notice that inside the contact zone delimited between $x=-90$ mm and $x=0$ mm the overall shape of $<U_{A}^{*}(x)>$ (left) is different compared to $<U_{B}^{*}(x)>$ (right). For A-types, the velocity is slightly decreasing from the inlet at $x=0$ mm down to a negative minimum near the leading edge at $x=-90$ mm and increases up to zero at the groove exit; for B-types, $<U_{B}^{*}(x)>$ is first decreasing down to a negative minimum around $x=-30$ mm in the first third of the contact zone length, and then increases up to positive values at the outlet of the contact zone around $x=-90$ mm. Therefore in A-grooves the absolute value of the fluid velocity, in a reference frame moving with the car, is higher than $V_{0}$ along the whole contact zone, while the fluid at the exit of B-grooves is sucked from the region located behind the tire, with absolute relative velocities smaller than $V_{0}$. It is remarkable that despite the lack of symmetry of C and D groove geometries, with respect to the vertical mid-plane of the tire, and within the limit of convergence of the ensemble averaging procedure, it is not possible to differentiate the flow inside the contact zone for A1 and A2, or for B1 and B2. The second major difference between A and B longitudinal grooves is the variability of velocity measurements. The standard deviation of velocity measurements appears to be much larger for B-type grooves than for A-types.

\begin{figure*}
   \centering
	\includegraphics[width=0.6\textwidth]{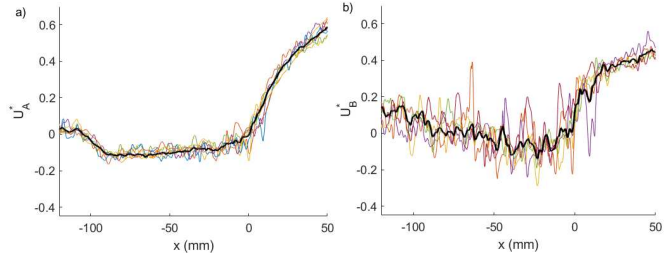}
	\caption{Instantaneous velocity profiles $U_{A}^{*}(x)$ (left) and $U_{B}^{*}(x)$ (right).}
	\label{fig18}
\end{figure*}

The physical nature of this spatial variability is highlighted when sets of instantaneous profiles $U_{A}^{*}(x)$ and $U_{B}^{*}(x)$ are plotted as in the Fig.\ref{fig18}. The small scale nature of velocity fluctuations is demonstrated in both cases. Nevertheless, while the magnitude of velocity peaks remains small for A-type grooves, higher local peaks are observed for B-type grooves. As suggested by the preliminary visualizations, these high amplitude peaks in B-type grooves may be linked to possible interactions between the flow inside C-grooves and B-type groove. The amplification of velocity fluctuations in B-type grooves is understandable if the flow in transverse grooves is investigated.

Results shown in Fig.\ref{fig18} were obtained by combining together all the data available in A1 and A2 grooves (resp. B1 and B2).

\subsection{Secondary vortical flows inside longitudinal grooves.}
\label{Vort}

The characterization of the secondary flows in a cross-section of a longitudinal groove is achieved through the analysis of the profiles of the measured transverse and vertical velocity components $V^{*}(x,y)$ and $W^{*}(x,y)$. In order to remove small scale fluctuations, $V^{*}$ and $W^{*}$ are spatially averaged over the length of the contact patch area : $V_{A1}^{*}(y)=\frac{1}{L_{c}}\int V^{*}(x,y)dx$ and $W_{A1}^{*}(y)=\frac{1}{L_{c}}\int W^{*}(x,y)dx$, where $L_{c}$ is the length of the contact patch area.

\begin{figure*}
	\centering
	\includegraphics[width=0.6\textwidth]{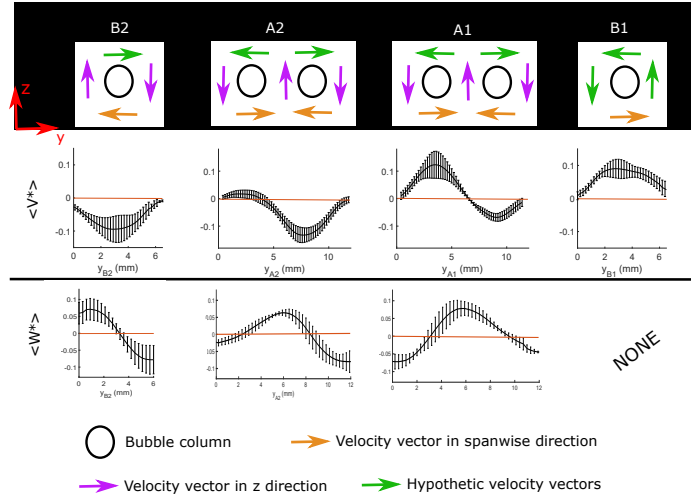}
	\caption{Scheme of longitudinal vortices and bubble columns inside grooves (up).
	V and W velocity profiles measured along the spanwise direction y (center and down)}
	\label{fig19}
\end{figure*}

Such velocity profiles measured in longitudinal grooves A1, A2, B1 and B2 are given in Fig.\ref{fig19}. The measured velocity profiles are in accordance with the sketch at the top of the figure showing a pair of counter-rotating vortices for A-type grooves, and one single vortex for B-type grooves.

For both A1 and A2 grooves, the evolution with $y^{*}$ of the sign of the measured transverse velocity $V^{*}$ is coherent with the direction of the flow in the lower part of the analytic vortex, near the ground and that is with an horizontal velocity directed towards the side walls of the groove. Moreover, the vertical measured velocity $W^{*}$ is also coherent with an upward motion of the liquid along the side edges of the grooves. Furthermore this is in accordance with the scheme proposed by Yeager in the case of two vortices. One needs to underline that, in the case of a vortex pair (A1 and A2 grooves), the actual strengths of the two counter rotating vortices seem to be unequals, while this is not the case for the analytical velocity field; it is likely that the actual vortices of a vortex pair are in fact not symmetrical, with the most intense vortex located near the central rib. 

The coherence of measurements with the proposed flow scheme is also fully achieved for smaller grooves B1 and B2 which were not notified in the paper of Yeager. It is remarkable to notice that the number of vortices appears to be proportional to the geometrical shape factor of the groove's cross-section, and coherent with the number of bubble columns identified with direct contact fluorescence visualizations. 

In conclusion, the secondary flow inside tire grooves is composed of two counter rotating vortices inside A-Type grooves and one single vortex in B-Type grooves (counter-rotating from one B groove to an other). Possible creation mechanisms of these vortices are proposed in Sec.\ref{Vortex}.

\subsection{Flow properties through C-type transverse grooves.}
\label{TC}

The analysis of the flow inside the transverse C-type grooves is of primary interest regarding the ability of a tire to drain water inside or outside the contact patch area. Following the same approach than for A and B grooves, we are focused on the $V^{*}$ velocity component parallel to the groove's direction. It is not possible to draw conclusions regarding the local flow structure inside such a C-groove since the local flow variability is too high and our spatial measurement resolution not high enough. Therefore for every individual transverse groove, $V^{*}$ is spatially averaged along its width and length, providing a global characteristic velocity $\overline{V_{C}^{*}}$. The longitudinal location $x_{C}$ of the center line of every C-groove, inside the contact patch area, is characterized by the dimensionless parameter $x_{C}/L_{c}$ varying in the range $0\%$ (inlet of the contact zone) to $100\%$.

\begin{figure*}
	\centering
	\includegraphics[width=0.6\textwidth]{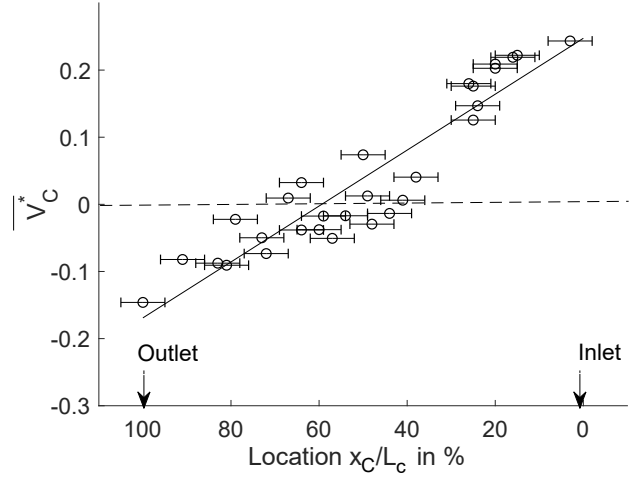}
	\caption{Spatially averaged velocity in C-type grooves vs location of the groove in the contact zone}
	\label{fig20}
\end{figure*}

The values of $\overline{V_{C}^{*}}$ are shown in Fig.\ref{fig20} as a function of $x_{C}/L_{c}$, for each transverse groove present inside the contact patch area for a given run and for each of the 16 independent runs. Despite the low number of statistical samples, the Fig.\ref{fig20} clearly shows a correlation between the characteristic velocity inside the groove and its longitudinal location $x_{C}$ along the contact zone; moreover the relationship between these parameters appears to be nearly linear. $\overline{V_{C}^{*}}$ is positive near the leading edge of the contact patch zone ($x_{C}/L_{c}=0\%$), which characterizes a global trend to eject water towards the outer part. This mechanism is at the origin of the elongated water filaments ejected above the liquid film, far from the external side of the tire. On the other hand, and more surprisingly, $\overline{V_{C}^{*}}$ is negative if the groove is near the downstream edge of the contact patch area : ($x_{C}/L_{c}=100\%$), with a flow radially directed towards the B-type groove geometrically connected to the transverse grooves under consideration.

In conclusion, in C-type grooves, an expulsion of the water is present at the beginning of the contact patch area and a suction effect appears at the end on this contact patch area. An explanation of this behaviour is proposed in Sec.\ref{TypeC}.

\section{Discussion.}

\subsection{Mass budget.}
\label{Mass}
In this section we explore the consequences of a mass budget performed inside the longitudinal grooves, having in mind to give a possible explanation of the different overall shapes of $<U^{*}_{A}>$ and $<U^{*}_{B}>$ noticed in the Fig.\ref{fig17}. With a reference frame moving at the car velocity, we define, in each longitudinal groove, a fixed control volume, with uniform cross-section dimensions (width w and height h). The first cross-section $S_{1}$ of the control volume is located at some distance downstream the contact zone inlet, e.g. at $x=-10$ mm, while the second zone $S_{2}$ is at $x=-90$ mm, before the outlet of the groove. The other boundaries of the control volume are the side walls, the top of the groove, and the ground. We assume that the measured velocity $<\overline{U^{*}(x)}>$ is uniform, representative at each $x$ cross-section of the bulk velocity. The mass budget in this control volume is averaged over all angular positions of the tire; therefore the instantaneous mass flux between a longitudinal groove and a transverse one occurring at random discrete positions of the transverse groove can be replaced by a steady mass flux $\dot{m}$ distributed along the left and right side walls of the longitudinal groove under consideration. From these conditions the mass budget applied to the control volume provides the following relation :
$-{\rho}wh(U_{2}-V_{0})+{\rho}wh(U_{1}-V_{0})+\dot{m}=0$ or:

\begin{eqnarray}
U_{2}=U_{1}+\frac{\dot{m}}{({\rho}wh)}
\label{mass}
\end{eqnarray}

For A-type grooves, we have seen, on visualization images, that mass flux exists with adjacent inclined D-type grooves.
The profile $<U_{A}^{*}(x)>$ in the Fig.\ref{fig17} (left) shows a small decrease of the velocity from $S_{1}$ down to $S_{2}$, indicating a small incoming (negative) mass flux through the side walls of the groove. Since the cross section of D-type grooves is small compared to the cross-section of A grooves, it is likely that the mass exchange through transverse D-grooves remains effectively small. 

On the other hand, for B-grooves, measurements in Fig.\ref{fig17} (right) clearly show that $U_{2}$-$U_{1}$ is positive, which could indicate an outcoming positive overall mass flux $\dot{m}$, mainly towards the C-grooves. The limit of this explanation is that the relative volume occupied by the bubble columns is higher in B-type than in A-type grooves, and the mass budget doesn't take into account this effect.

\subsection{Self-similarity of the flow.}
\label{Similarity}

Direct visualizations made by Todoroff et al.2018\cite{todoroff2018mechanisms} using fluorescein show a decrease of the contact patch length for increasing values of $V_{0}$. In this section we analyse the influence of $V_{0}$ on velocity profiles inside longitudinal grooves, and the possibility to obtain self-similar profiles using a suitable rescaling of the longitudinal coordinate $x$. 

\begin{figure*}
	\centering
	\includegraphics[width=1\textwidth]{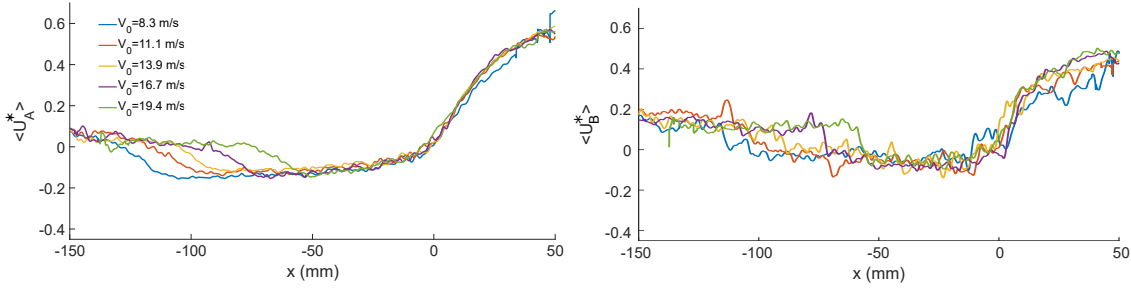}
	\caption{Velocity profiles in A-grooves (left) and B (right) grooves for different car speeds $V_{0}$.}
	\label{fig21}
\end{figure*}

$<U_{A}^{*}>$ and $<U_{B}^{*}>$ ensemble averaged velocity profiles obtained from r-PIV measurements as a function of $V_{0}$, changing from $8.3$ m.s$^{-1}$ up to $19.4$ m.s$^{-1}$, are shown in the Fig.\ref{fig21}, respectively for A-type (left) and B-type (right) grooves. For $<U_{A}^{*}>$ the curves (left) show that the main effect of an increase of $V_{0}$ is to progressively induce a reduction of the length $L_{p}$ of the plateau with negative velocities.

\begin{figure*}
	\centering
	\includegraphics[width=0.5\textwidth]{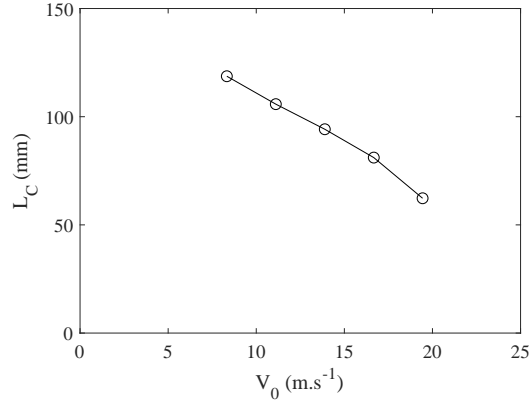}
	\caption{Evolution of contact patch length $L_{C}$ with car speed $V_{0}$}
	\label{fig22}
\end{figure*}

Inside B-grooves (right) the different profiles $<U_{b}^{*}>$ don't collapse perfectly, due to the highly spatially intermittent characteristic of the profiles linked to the low number of statistical samples inducing a low convergence. We define a characteristic length scale $L_{p}$ as the distance between the inlet $x=0$ mm and the point where the leading edge of the profile reaches half its maximum negative value. The overall decrease of $L_{p}$ for increasing $V_{0}$, depicted in the Fig.\ref{fig22}, is coherent with the measurements of the contact patch length $L_{c}$ deduced from fluorescein visualizations by Todoroff et al. 2018\cite{todoroff2018mechanisms}.

\begin{figure*}
	\centering
	\includegraphics[width=1\textwidth]{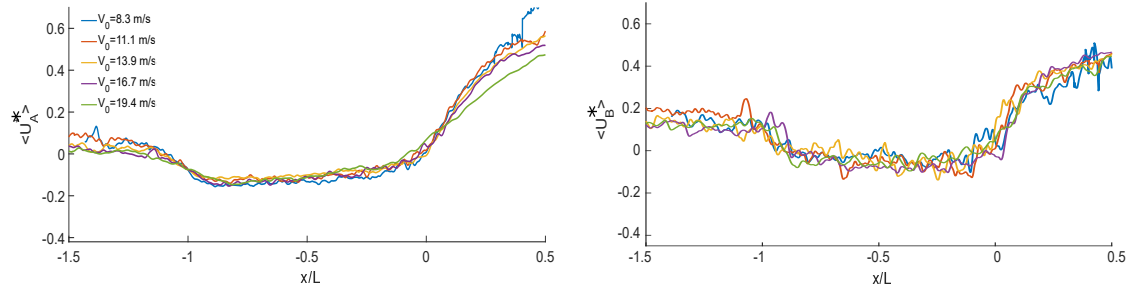}
	\caption{Self-similarity of velocity profiles in longitudinal grooves for increasing $V_{0}$ in A-grooves (left) and B (right)}
	\label{fig23}
\end{figure*}

Velocity profiles in the Fig.\ref{fig23} show the self similar behaviour of the flow inside longitudinal grooves provided by a normalisation of the $x$ axis by $L_{p}$ and a normalisation of the velocity by $V_{0}$. The collapse of the different curves is very good inside the plateau for A-type grooves, and satisfactory for B-type grooves if we consider the limited number of B-C groove hydrodynamic interactions covered by the available database. 

Therefore the behaviour of the velocity field inside longitudinal grooves can be considered as self similar for increasing vehicle speeds when using suitable length and velocity scales. Cabut et al. \cite{cabut2019} have shown that the displacement of the tire induces, in front of the central rib, a sharp velocity front convected at the car speed; the distance between the rib and this velocity front increases in a non-linear manner with the car speed, while in this part of the flow, the characteristic velocity of the water film increases linearly for moderate car speeds, with some non-linear behaviour for higher velocities when the tire tends to an hydroplaning situation. As a conclusion, we conclude that both external (in front of the tire) and internal (inside grooves) zones of the flow show self-similar structures for increasing velocities, using the length of the contact patch area as a characteristic length. 

\subsection{Vortex creation mechanism.}
\label{Vortex}

A detailed analysis of the creation mechanisms of vortices and bubbles, and their possible interactions, is far beyond the scope of the present work; some elementary possible mechanisms are now discussed.
The topic of vortices inside grooves of a tire rolling on a wet road is to our knowledge extremely poor in the literature, except the pioneering work of Yeager 1974\cite{yeager1974tire} previously mentioned. It is here difficult to know the exact source of the vortices creation in this highly complex free-surface flow. Some works are focused on the analysis of the creation of horseshoe vortices for a partially submerged obstacle in a steady free surface flow (Launay 2017 \cite{launay2017experimental} and Chou and Chao 2000 \cite{chou2000branching}). However, in the tire case we are facing a case where the flow is more complex with unsteady behaviour, and a non symmetrical moving object. Inside tire grooves, counter rotating vortices have been observed once in the literature in Croner's work in 2014\cite{croner2014etude}. These vortices were observed from results obtained with RANS simulations of the air flow around a rolling tire. Therefore these vortices can appear for a single phase air flow inside the grooves of a tire with a slightly different geometry.

\begin{figure*}
    \centering
    \includegraphics[width=0.7\textwidth]{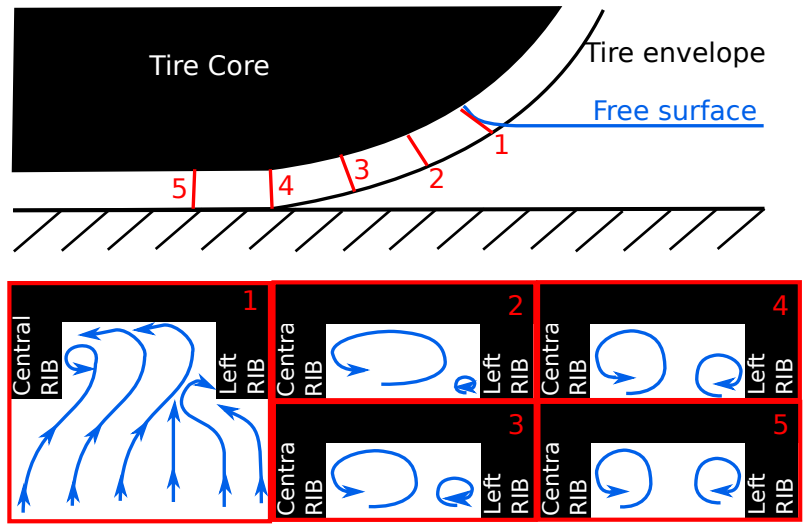}
    \caption{Sketch of the tire case with the vortex creation evolution between cross-section 1 and 5.}
    \label{fig25}
\end{figure*}

Therefore, we can propose here two different mechanisms which can explain the creation of these vortices with a single phase flow as well as a free surface flow. The first creation mechanism could be linked to the flow around the sharp edge of the tire rib. This effect is similar to the one observed for delta wings in aerodynamic. These instabilities have been extensively studied in the literature by Miau et al. 1995 \cite{miau1995flow} who studied the vortex formation at the leading edge of a delta wing with an incidence angle of $10^{\circ}$. The explanation of the formation of these vortices were discussed by Gad-el-Hak et al. 1985 \cite{gad1985discrete} based on flow visualisations. They observed the classical large vortices originated from smaller merged vortices created at the leading edge of the delta wing. In the tire case, proposal consists in the creation of the vortices at the RIB sharp angles at the beginning of the contact patch area as sketched in Fig.\ref{fig25}. A vortex is further convected, though the contact patch area, with the water flow, in the $x$ direction. 

\begin{figure*}
    \centering
    \includegraphics[width=0.8\textwidth]{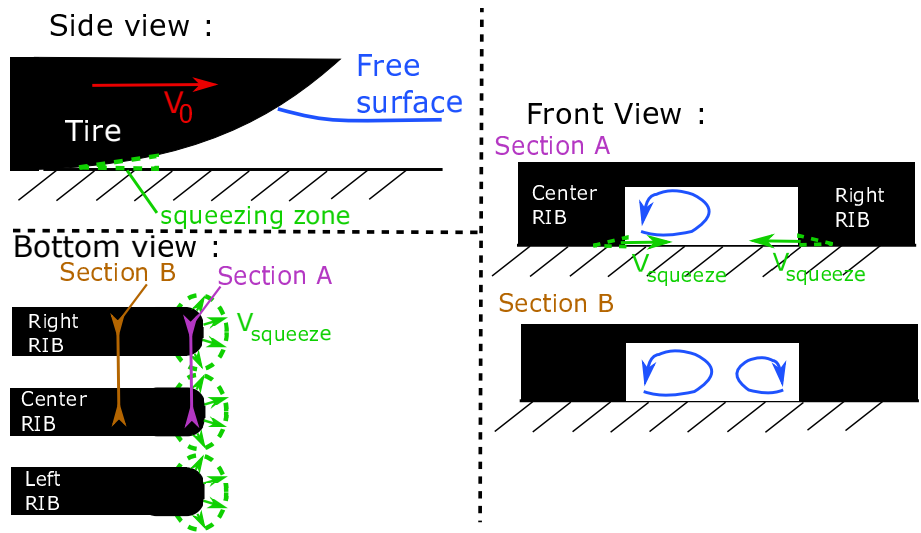}
    \caption{Squeezing effect and creation of a second vortex in A-type grooves.}
    \label{fig26}
\end{figure*}

A second vortex creation mechanism can be due to the squeezing effect at the entry of the contact patch area. Near the contact patch area, the load exerted by the tire on a thin layer of water will generate a squeezing which can allow the creation of two vortices with a transverse injection of water in the lower part of the groove on each side as sketched in the Fig.\ref{fig26}. 

Furthermore, the creation of vortices is conditioned by the complex flow situation in the puddle zone and close to the groove inlet.

A last point to notice is that the bubble column, that could be trapped at a vortex center all along a groove, is a matter of interesting future work. It is not clear if they are due to air entrainment from the free surface at the groove inlet.
  
As mentioned previously, the precise analysis of the creation mechanism  involved for these vortices if far beyond the scope of this work. In this section, we have proposed two possible mechanisms that can create vortices inside tire grooves. Further studies and analysis should be done in order to exactly understand how the vortices appear.

\subsection{Deformation of C-type grooves.}
\label{TypeC}

In the transverse grooves, observations have been made on the water velocity parallel to the groove directions for PCY4 C-type grooves (Section.\ref{TC}). It appears that the velocity inside the tire groove is highly correlated with its position in the contact patch area. This phenomenon could be explained by the deformation of the rubber block between grooves when the tire load exerts a compression on it. In the contact patch area the pressure load is of a complex distribution as shown by Tielking and Roberts 1987 \cite{tielking1987tire} based on different experimental sources. However, at the shoulder, the pressure distribution in the $x$ direction can be approximated by a parabolic curve, as used by Heinrich and Kluppel 2008 \cite{heinrich2008rubber} for their analysis of the friction coefficient during braking as $p(x)=4.p_{m}.x/L_{c}.\left(1-x/L_{c}\right)$, with $p_{m}$ the averaged pressure and $L_{c}$ the length of the contact patch area.

\begin{figure*}
    \centering
    \includegraphics[width=0.8\textwidth]{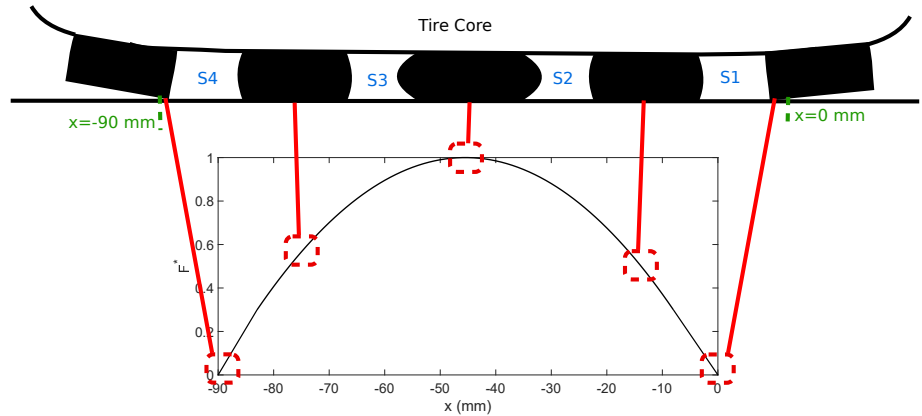}
    \caption{Schematic deformation of C- type grooves under mechanical load.}
    \label{fig27}
\end{figure*}

Thus, the total load exerted on a rubber block at the shoulder can be written as : $F=w_{b}.\int_{x_{min}}^{x_{max}}p(x)dx$; where $w_{b}$ is the width of the rubber block and $x_{min}$ and $x_{max}$ are the limits of the block in the $x$ direction. With the exerted load the compression of the rubber block appears and a deformation with it. A sketch Fig.\ref{fig27} illustrates the deformation of the rubber block as a function of the load on the block. The normalised load is calculated as $F^{*}=F/max(F)$ and is represented considering a rubber block size of $\Delta_{b}=x_{max}-x_{min}=15$ mm which is approximately the size of the rubber block at the shoulder of the PCY4 tire.

During the compression, the rubber block is subject to a Poisson's ratio expansion. This deformation of rubber block depends on its elasticity properties and were extensively studied in the literature as by Gent and Lindley 1959 \cite{gent1959compression} for which the influence of a corrected Young Modulus function of the shape ratio of the rubber block was discussed. This deformation of the rubber block can not be precisely calculated here as we don't know the exact composition and properties of the tire rubber. However, the shape of the deformation of a rubber block as sketched in the Fig.\ref{fig27}, allows us to deduce the reduction of the groove section between the beginning of the contact patch area and the center ($S1>S2$). It follows the growth of the groove section between the center of the contact patch area and the end ($S3<S4$). The cross section surface of the groove is reduced at the beginning of the contact patch area which will push the water out of the groove. At the end of the contact patch area, the relaxation of the rubber block generates a reopening of the tire groove which will induce a suction.

\section{Conclusions.}
\label{Conc}

In this paper, we demonstrated, for the first time, the feasibility of velocity measurements inside some groove flows of a real rolling tire using the r-PIV method, with the help of more classical contrast fluorescence visualizations of the contact patch area. The analysis of both visualizations and PIV raw images highlights the two-phase nature of elongated structures present in some specific parts of the grooves. This two-phase flow is shown to produce an extinction of the light re-emitted by fluorescent particles located in the upper part of the grooves. This, therefore, operates inside the illumination volume a selection of the reemitted fluorescent light.

The analysis of the flow structure is based on one-shot measurements repeated over sets of independent runs in order to study the variability of the flow. This analysis combines both ensemble averaging and spatial averaging in order to extract some characteristic velocities inside the grooves. 

Measurements have shown that the primary flow inside longitudinal grooves is not the same along the contact zone length when considering either A or B-type grooves. The differences are presumed to be related to hydrodynamic interactions with either C or D-type grooves. This technique also allows us to prove the existence, and to analyse, the vortical structure of the flow in these longitudinal grooves, the link between the number of vortices developed and the aspect ratio of the grooves.
The precision of the measurements is sufficient to also determine the evolution of the water flow through transverse C-grooves determining the dynamic evolution of this flow with the position of the groove in the contact patch area. Up to now the characterization of the flow in transverse grooves relies on a global quantity spatially integrated over the groove's area; a more comprehensive analysis of the local hydrodynamic interactions between longitudinal and transverse grooves should be performed with an higher spatial resolution. 
Finally results given in the present paper are focused on the case of a new tire with a relatively high water height. From an experimental point of view, the investigation of cases with smaller water height for a worn tire remains a challenging objective; in that case the modifications of the geometric shape factor of groove's cross-section is likely to modify the creation and hydrodynamic of the longitudinal vortical structures. 

\newpage

\section*{Acknowledgments :}
This work was supported by BPI France (grant n$^{\circ}$ DOS0051329/00) and AURA Region (grant n$^{\circ}$ 16 015011 01) in the framework of the HydroSafeTire project.

\end{document}